\documentclass[12pt]{iopart}

\usepackage{epsf}
\usepackage{graphics}
\usepackage{graphicx}

\newcommand{\Zwicknagl}{partially localized}
\newcommand{\Norman}{fully itinerant}

\begin{document}

\title[The Fermi surface of UPt$_3$]{The 
Fermi surface and f-valence electron count of UPt$_3$}

\author{
G J McMullan$^1$, 
P M C Rourke$^2$, 
M R Norman$^3$, 
A D Huxley$^4$, 
N Doiron-Leyraud$^5$, 
J Flouquet$^6$, 
G G Lonzarich$^7$, 
A  McCollam$^2$, 
and 
S R Julian$^{2,8}$. 
}

\address{$^1$ MRC Laboratory of Molecular Biology,
Hills Road, Cambridge, CB2 0QH, UK}
\address{$^2$ Department of Physics, University of Toronto, 
Toronto, ON M5S 1A7, Canada}
\address{$^3$ Materials Science Division, Argonne National Laboratory,
Argonne, IL 60439, USA}
\address{$^4$ School of Physics, James Clerk Maxwell Building, 
King's Buildings, Mayfield Road, Edinburgh EH9 3JZ, UK}
\address{$^5$ D\'epartement de Physique, Universit\'e de Sherbrooke, 
Sherbrooke, PQ J1K 2R1, Canada}
\address{$^6$ D\'epartement de Recherche Fondamentale sur la Mati\`ere 
Condens\'ee, SPSMS, CEA/Grenoble, 17 rue des Martyrs, 38054 Grenoble cedex 9, 
France}
\address{$^7$ Cavendish Laboratory, University of Cambridge, Madingley Road, 
Cambridge, CB3 OHE, UK}
\address{$^8$ Author to whom any correspondence should be addressed.}
\ead{sjulian@physics.utoronto.ca}

\date{\today}

\begin{abstract}
Combining old and new de Haas-van Alphen (dHvA) and magnetoresistance 
data, we arrive at a detailed picture of the Fermi surface of the heavy
fermion superconductor UPt\(_3\).
Our work was partially motivated by a new proposal that two 5f valence electrons per
formula unit in UPt\(_3\) are localized by correlation effects---agreement with
previous dHvA measurements of the Fermi surface was invoked in its support. 
Comprehensive comparison with our new observations shows that 
this `partially localized' model fails to predict the existence of a major sheet of the Fermi 
surface, and is therefore less compatible with experiment than the originally proposed 
`fully itinerant' model of the electronic structure of UPt\(_3\).
In support of this conclusion, we offer a more complete analysis of the
fully itinerant band structure calculation, where we find a number of
previously unrecognized extremal orbits on the Fermi surface.
\end{abstract}
\submitto{\NJP}


\maketitle

\section{Introduction} 

The heavy fermion superconductor UPt$_3$ is regarded as an
archetypal strongly correlated Fermi liquid.  
At high temperatures, the
uranium 5f electrons show local moment Curie--Weiss behaviour, and
they strongly scatter the conduction electrons to give a large
resistivity.   Below a few degrees Kelvin, however, the scattering becomes
coherent, and quasiparticle bands populated by massive charged
fermions form.   Indirect evidence for the
existence of heavy quasiparticles comes from thermodynamic and
transport properties, for example, the enormous linear coefficient of the
specific heat, $\gamma \sim 420$ mJ\,(mole U)$^{-1}\,$K$^{-2}$  
\cite{Frings83,Stewart84}, 
a similarly huge (and anisotropic) Pauli-like susceptibility,
$\chi_c \sim 50 \times 10^{-9}$ and 
$\chi_{a,b} \sim 100 \times 10^{-9}$ m$^3$(mole U)$^{-1}$ for fields along 
the $c$-axis and in the basal plane, respectively  
\cite{Frings83}, and a very large $T^2$
coefficient of the resisitivity, $A \sim 0.49\,  
\mu\Omega\,$cm$\,$K$^{-2}$ \cite{Taillefer88a}.  
More direct
evidence comes from optical conductivity \cite{Marabelli87} and from the
observation of a Fermi liquid contribution to the dynamical magnetic
susceptibility as measured by inelastic neutron scattering \cite{Bernhoeft95}.

\begin{figure}
\begin{center}
\leavevmode
\epsfysize 19cm
\epsfbox{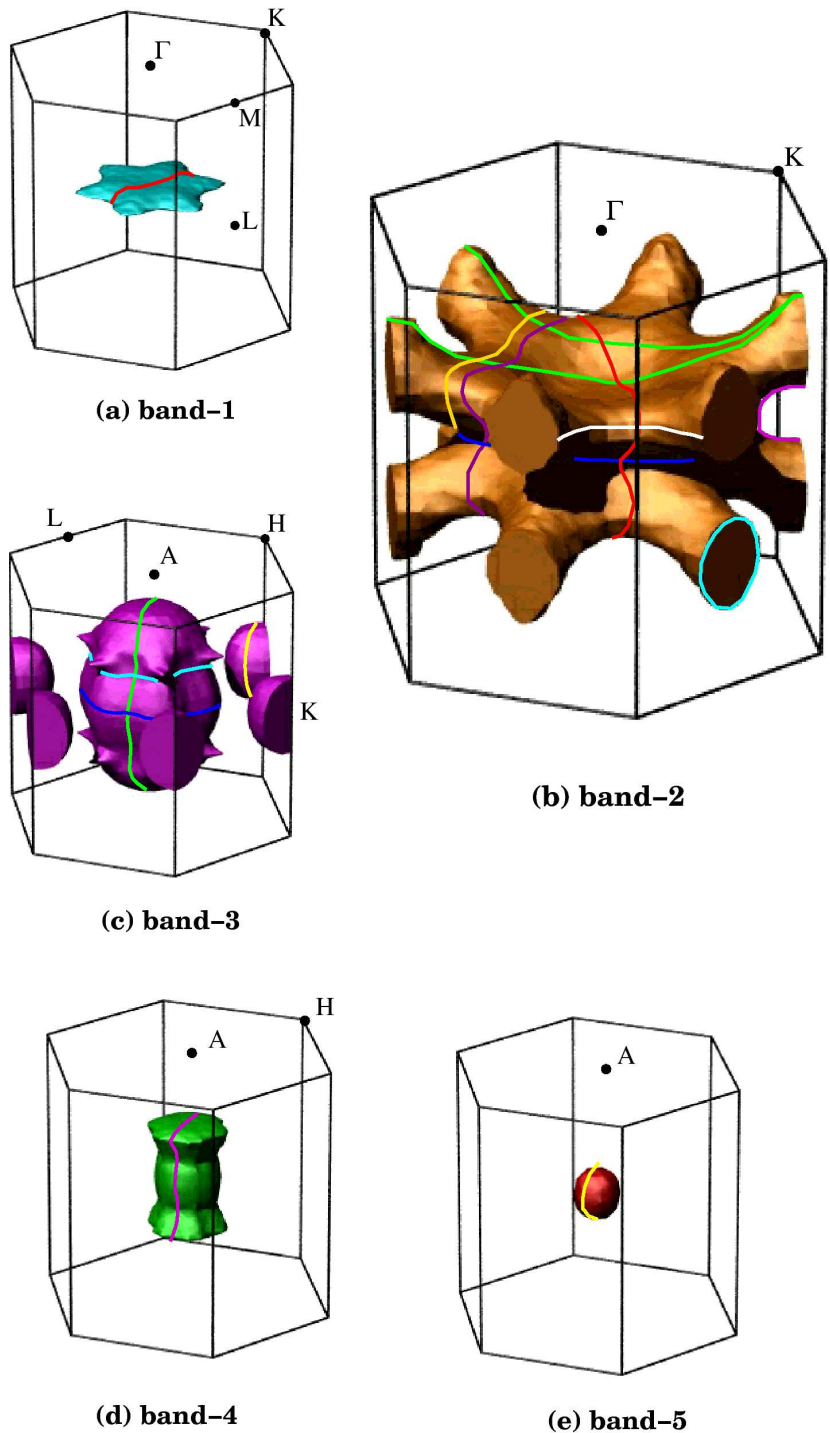}
\end{center}
\caption{Fermi surfaces generated from the \Norman\ model,
which assumes all of the uranium 5f-electrons
are included in the Fermi volume.
The band 1 and 2 surfaces are hole surfaces;
the band 3, 4 and 5 surfaces are electron surfaces.
All of the simple extremal orbits on each surface are shown. 
Table \ref{tab-orbits} gives the correspondence between 
these lines and predicted dHvA frequencies of figure \ref{Freqs_vs_FS}. 
On the band 2 surface, the two green lines running across the 
top indicate possible `open orbits'. 
Note also that subsequent plots are colour coded 
such that dHvA frequencies thought to arise from 
a given band are shown in the same colour as the Fermi surface in this 
figure (e.g.\ band 2 frequencies will be plotted in brown, etc.) 
}
\label{valence_FS}
\end{figure}

Taillefer and Lonzarich provided the most direct evidence for the existence 
of the heavy Fermi liquid, 
with their 
measurements of quantum oscillatory magnetization---the 
de Haas-van Alphen (dHvA) effect  
\cite{Taillefer_87,Taillefer_88}. 
They observed at least five distinct extremal orbits of the Fermi surface, 
with quasiparticle masses of up to 100 $m_e$. 
All of the observed surfaces 
could be interpreted by comparison with 
relativistic band structure calculations
\cite{Norman88a},  
which predict the multi-sheeted 
Fermi surface shown in figure \ref{valence_FS}.  
The observed quasiparticle masses were enhanced by 
between 10 and 20 in comparison 
with the calculated band mass. 
Increasing the calculated contribution to \( \gamma \) of each Fermi surface sheet 
by the ratio of the measured to the 
calculated effective mass 
gave good agreement with the total measured linear coefficient 
of the specific heat: that 
is, these five sheets of the Fermi surface were found to 
account within error for the observed 
low temperature linear specific heat coefficient \cite{Taillefer_88}.  
These dHvA studies of UPt\(_3\), and their interpretation, 
were very influential in the development of 
our present understanding of heavy fermion systems, 
so it is essential to establish them unambiguously, the 
more so because 
the interpretation of the dHvA data has recently been 
called into question by an alternative proposal that 
two of the uranium \(5f\)-electrons are localized 
in UPt\(_3\) \cite{Zwicknagl}. 
This proposal has the attractive feature that it 
offers a specific mechanism for producing quasiparticle 
mass enhancements: 
the exchange interaction between 
the itinerant and localized f-electrons. 
However, this explanation of the mass enhancement is not 
unique: in addition to the well-established Kondo-lattice mechanism~\cite{Auerbach,Millis} which is believed to explain heavy masses in itinerant 1f cerium heavy fermion compounds, such as CeRu\(_2\)Si\(_2\)~\cite{Aoki}, one can get large mass enhancements from coupling of conduction electrons to their own collective spin-fluctuations~\cite{Lonzarich85,NormanPRL87,Moriya} as in MnSi (see e.g.~\cite{Lonzarich88}). Thus, the argument for adopting the `\Zwicknagl' 
model rests on the claim that it gives a better description of 
of the Fermi surface than the original `\Norman' model. 
One of the central goals of this paper is to critically examine this claim. 
 
There are other reasons why it is important that we understand the 
electronic structure of UPt\(_3\). 
UPt\(_3\) is the archetypal multi-component superconductor 
(for reviews see references \cite{Joynt_02,Sauls_94}),  
showing three distinct superconducting phases below \(T_c \simeq 0.5 {\rm\ K}\).
It is crucial that the correct Fermi surface be employed 
in attempting to model thermodynamic measurements in the 
superconducting state (see e.g.\ \cite{Norman96}) 
and that the correct elementary excitations are 
used in trying to understand the superconducting pairing mechanism: 
in particular, the symmetry of the pair state will be sensitive to the 
details of the electronic structure of UPt\(_3\). 

From a theory perspective, the nature of the Fermi surface comes to
the heart of the debate concerning the behavior of f-electrons in condensed
matter systems.  
Are they localized, itinerant, or is a `dual' model (partially
localized and partially itinerant) more appropriate?
For instance, in UPd\(_3\), a local-f model gives a good
description of the Fermi surface \cite{Norman87}, whereas it gives a very poor
description of the Fermi surface of UPt\(_3\) \cite{Norman88a}.  
Since UPt\(_3\) sits near the borderline between local and itinerant behavior, it is
not a priori clear whether a `dual' or a `fully itinerant' picture is more appropriate.
 
Finally, there is obvious interest in improving the accuracy of 
band structure calculations for correlated electron systems,
and UPt\(_3\) provides a good test case since it sits very nicely between 
magnetic d-electron metals, which have fairly broad bands and 
for which local density approximation (LDA) calculations seem to work well, and 
more strongly correlated metals, such as sodium cobaltate, in which it is 
claimed that LDA calculations give an incorrect electronic structure 
\cite{Yang05}.

\section{Theoretical models of the Fermi surface}

In this paper, the aspect of electronic structure that we focus on 
is the Fermi surface. 
Using the Onsager relation (see below) 
a dHvA 
oscillation frequency 
can be convered into a 
cross-sectional area of 
a sheet of the Fermi surface that 
is `extremal' (a maximum or minimum) in the plane 
perpendicular to the direction of the 
applied magnetic field.  
By rotating the crystal 
in the field the 
angle dependence of these extremal areas 
can be followed.
It is sometimes possible to derive 
the shape of 
the Fermi surface 
directly 
from such data, 
but in practice it is usually 
necessary to interpret the data 
by comparison with the 
Fermi surfaces predicted by 
band structure calculations. 
Taillefer and Lonzarich \cite{Taillefer_88} used a 
Fermi surface generated by Norman \etal  \cite{Norman88a}, 
who calculated
the band structure 
assuming that all the uranium 5f electrons are
itinerant (i.e. that the Fermi volume includes these electrons).  
This Fermi surface, 
shown in 
figure  \ref{valence_FS} 
together with the path in \(k\)-space 
of several of the allowed extremal orbits, was in general agreement with earlier calculations~\cite{Oguchi,Wang87,Norman88b}, but there are differences in detail. 
Five bands cross the Fermi energy, 
and the five panels of figure \ref{valence_FS}  
show the resulting sheets of the Fermi surface: 
(band 1) a disc-shaped hole surface centred on the
$A$ point of the Brillouin zone; 
(band 2) a multi-connected  hole 
surface centred on $A$ with arms branching out to reach the zone
boundary between $L$ and $M$; 
(band 3) a large 
$\Gamma$-centred electron surface with subsidiary electron 
pockets centred at $K$; 
(band 4) and (band 5) two smaller $\Gamma$-centred electron surfaces.

\begin{table}
\caption{Comparison of notation used for Fermi surfaces of UPt\(_3\): 
the `band number' is the notation used by Norman \etal \cite{Norman88a} and in this paper. 
The other references are:  
Joynt and Taillefer \cite{Joynt_02} (a review article on the
superconductivity); Taillefer and Lonzarich \cite{Taillefer_88};
Kimura \etal \cite{Kimura98}; Zwicknagl \etal \cite{Zwicknagl}.
`Colour' is the colour scheme used to represent a given band on plots of 
dHvA versus angle.
}
\label{table:nomenclature}
\begin{indented} \item[]
\begin{tabular}{|c|c|c|c|c|c|}
\hline
Band number    &  Joynt  \& &  Taillefer  \& & Kimura & Zwicknagl & Colour \\
               &  Taillefer        &  Lonzarich     &  \etal & \etal   &   \\
\hline
Band 1  &   `Starfish' & Band 5  &  Band 35 & Band 1 (Z1)  & Turquoise 
\\ \hline
Band 2  &   `Octopus' & Band 4  & Band 36  &  --- & Brown \\
        &   &         &          &  & 
\\ \hline
Band 3  &   `Oyster \& & Band 3 & Band 37 & Band 2 (Z3) & Purple \\
        &   urchins'  &          &          &   & 
\\ \hline
Band 4  &   `Mussel'  & Band 2 & Band 38 & --- & Green 
\\ \hline
Band 5  &   `Pearl'  & Band 1     & Band 39 & Band 2 (Z5) & Red
\\ \hline
\end{tabular}
\end{indented}
\end{table}

The notation describing the Fermi surface and dHvA effect in UPt\(_3\) has 
become rather complicated over the years. 
In this paper, 
in keeping with the original notation of Norman \etal \cite{Norman88a},  
we are referring to the  five bands that cross the Fermi surface as 
bands 1 through 5, while particular predicted orbits 
will be labelled by giving the 
centre of the orbit and the band number.
Thus, for example, 
A-1 is an extremal orbit on band 1, with the orbit centred on the A 
point in the Brillouin zone (A-1 is indicated by the red line in figure 
\ref{valence_FS}(a)).  
Note, however, that experimentially observed (as opposed to theoretically 
predicted) dHvA frequencies will 
be referred to by lower-case Greek letters, 
following the notation of references \cite{Taillefer_88} and \cite{Kimura95,Kimura96,Kimura98}. 
Different band numbering has been used by different authors, 
so in table \ref{table:nomenclature} 
we tabulate the various labelling schemes that have been used, 
to ease comparison with the published literature. 

In \cite{Taillefer_88}, agreement between the
observed extremal orbits and this calculated Fermi
surface was based on a number of points of detail, but rigorous
comparison was not possible because quantum oscillations from most Fermi
surface sheets were only seen with the field close to the $a$-axis 
(which corresponds to the \(\Gamma-K \) direction in \(k\)-space), and
{\em no}
oscillations were observed with the field  along the $c$-axis (field parallel to 
\(\Gamma-A\)).   Moreover,
some predicted sheets were not observed at all.
Many of these gaps were filled by Kimura \etal
\cite{Kimura95,Kimura96,Kimura98},
who  observed quantum oscillations throughout all three 
major symmetry planes and along  the
$c$-axis.

\begin{figure}
\begin{center}
\leavevmode
\epsfxsize 10cm
\epsfbox{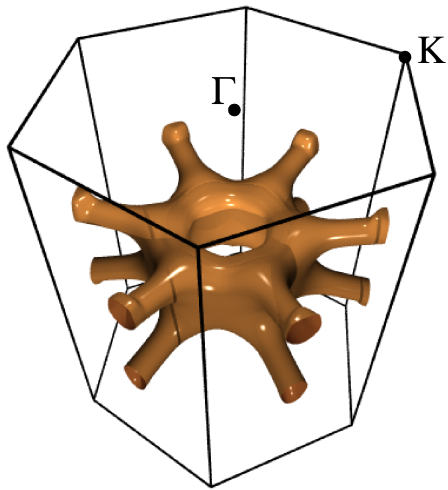} 
\end{center}
\caption{Cartoon of the modified band 2 surface proposed by Kimura \etal\ \cite{Kimura95}.  
The hole through the centre of the surface eliminates the open orbits of figure \ref{valence_FS}(b). }
\label{kimura_FS}
\end{figure}

It has twice been proposed 
that 
the Fermi surface shown in figure \ref{valence_FS} 
is wrong.
The first proposed change, 
due to Kimura \etal \cite{Kimura95}, is illustrated in figure \ref{kimura_FS},
where the band 2 surface --  
in the \Norman\ model a large A-centred hole surface 
with twelve arms extending to the Brillouin zone boundary --
has become a large {\em torus} with twelve arms extending to the zone boundary.
In this torus geometry there are no `open orbits' on the Fermi surface.
(In figure \ref{valence_FS}(b), the open orbits are shown
as green lines running across the top of the Fermi surface, spanning
the Brillouin zone.)
Such orbits are
important in magnetoresistance, because 
cyclotron motion of quasiparticles
arising from the Lorentz force
is prevented on open orbits 
when the applied magnetic field is perpendicular to the orbit:
in effect, the quasiparticles on the open orbits,
rather than going in circles in both \(k\)-space, and real space,
run along the top of the surface in \(k\)-space and thus
move in roughly a straight line in real space.
In copper, for example, this produces very pronounced minima in the 
high field magnetoresistance
when the applied field is perpendicular to the open orbit
(see e.g. \cite{PippardBook}),
and indeed
Kimura \etal
eventually reverted to the original topology of this surface as shown in 
figure 1(b) \cite{Kimura98},
largely because 
the magnetoresistance shows such angular dependent minima, 
as discussed in more detail below.

\begin{figure}
\begin{center}
\leavevmode
\epsfysize 3in
\epsfbox{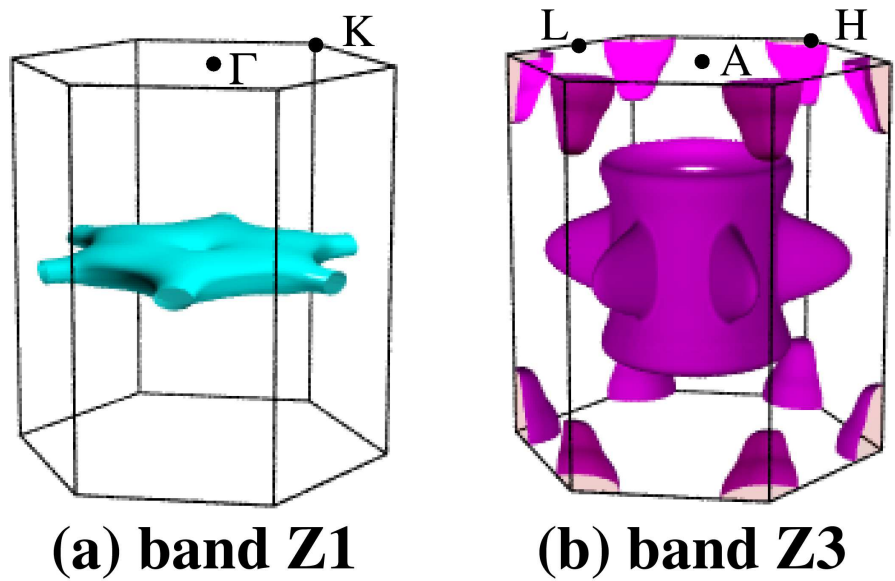}
\end{center}
\caption{Cartoon representation of the major Fermi surface sheets proposed by 
Zwicknagl \etal \cite{Zwicknagl}.  
The `Z1' surface has open orbits similar to those of band 2 of
the \Norman\ model.
The `Z3' surface has a large central electron surface plus small
H-centred electron pockets. 
A small electron surface (Z5), similar to the band 5 surface of the
\Norman\ model, is not shown.}
\label{Zwicknagl_FS}
\end{figure}

The existence of these open orbits is relevant to the
more recent, and more radical, proposal by 
Zwicknagl \etal \cite{Zwicknagl}
of a dual model for 
UPt\(_3\), in which 
two of the three uranium 5f-electrons
per formula unit are localized in the
sense that they do not contribute to the Fermi volume. 
We will refer to this as the `\Zwicknagl'\ model,
in contrast to the original `\Norman'\ model.
Figure \ref{Zwicknagl_FS} shows the 
Fermi surface for this model. 
It has two large surfaces, which 
we label Z1 and Z3 as they seem to 
correspond most closely to the band 1 and band 3 
surfaces of the \Norman\ model, 
plus a small \( \Gamma\)-centred electron pocket (not shown).

The band Z1 Fermi surface
is similar in shape to the band 1 surface of figure \ref{valence_FS}(a),
however, in the \Zwicknagl\ model it
spans the Brillouin zone
at the L-points,
producing open orbits when the applied field is
parallel to
the \(a\)-axis.

The large \(\Gamma\)-centred electron surface,
arising from band Z3,
is somewhat larger than,
and somewhat different in shape from,
the large band 3 surface of the \Norman\ calculation.

The correspondence between the bands should not be taken too literally, 
as can be understood 
by examining degeneracies of the bands that arise within the
hexagonal close packed crystal structure.  
(The actual crystal structure of UPt\(_3\) is hexagonal close packed with a slight trigonal distortion \cite{Walko}, but this distortion was not included in the band structure calculations and should have only a tiny effect.) In the hexagonal close packed structure each primitive
cell contains two formula units, and thus the bands are related
in pairs.  
These pairs would be degenerate in the A-H-L plane,
but this degeneracy is mostly lifted by spin-orbit coupling, leaving
only a degeneracy along the A-L lines.  
It can be seen that in the A--H--L plane the Fermi surfaces of bands 1 and 2 of the \Norman\ model are nearly degenerate, and they are degenerate along the A--L line \cite{}; similarly, we see that in the A--H--L plane the H-centred ellipsoids of band Z3 are nearly degenerate with the Z1 surface, implying that these H-centred ellipsoids are the analogue of band 2 of the \Norman\ model; it is only the large $\Gamma$-centred surface of band Z3 that corresponds to band 3, and we also infer that the analogue of band 4 of the \Norman\ model is the small $\Gamma$ surface of the \Zwicknagl\ model (not shown in figure \ref{Zwicknagl_FS}).
Despite this minor confusion, we will label
the H ellipsoids as belonging to `Z3', and the smaller $\Gamma$ surface
as belonging to `Z5', as this facilitates comparison of the two models
in regards to the dHvA data.

\begin{figure}
\begin{center}
\leavevmode
\epsfxsize 15.0 cm
\epsfbox{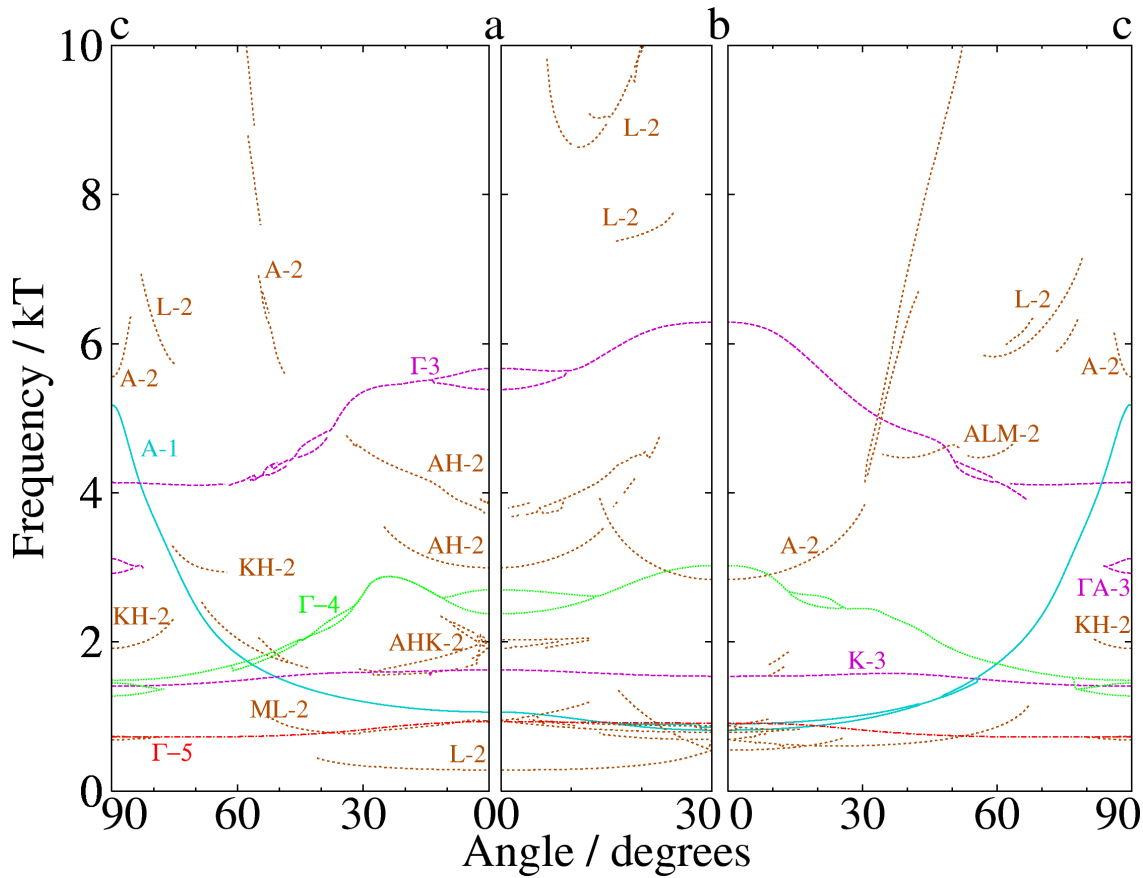}
\end{center}
\caption{
Predicted variation of the dHvA frequency components in
the \Norman\ model.  The labels give the orbit centre followed by the band number (1--5). 
Table \ref{tab-orbits} gives the corresponding orbits in figure \ref{valence_FS}. 
At the \(c\)-axis the applied field is along the hexagonal axis, at \( a \)  
it is parallel to \(\Gamma\)-K (see figure \ref{valence_FS}), whereas at \( b\) 
it is parallel to \(\Gamma\)-M. 
}
\label{Freqs_vs_FS}
\end{figure}

\begin{figure}
\begin{center}
\leavevmode
\epsfxsize 12.0 cm
\epsfbox{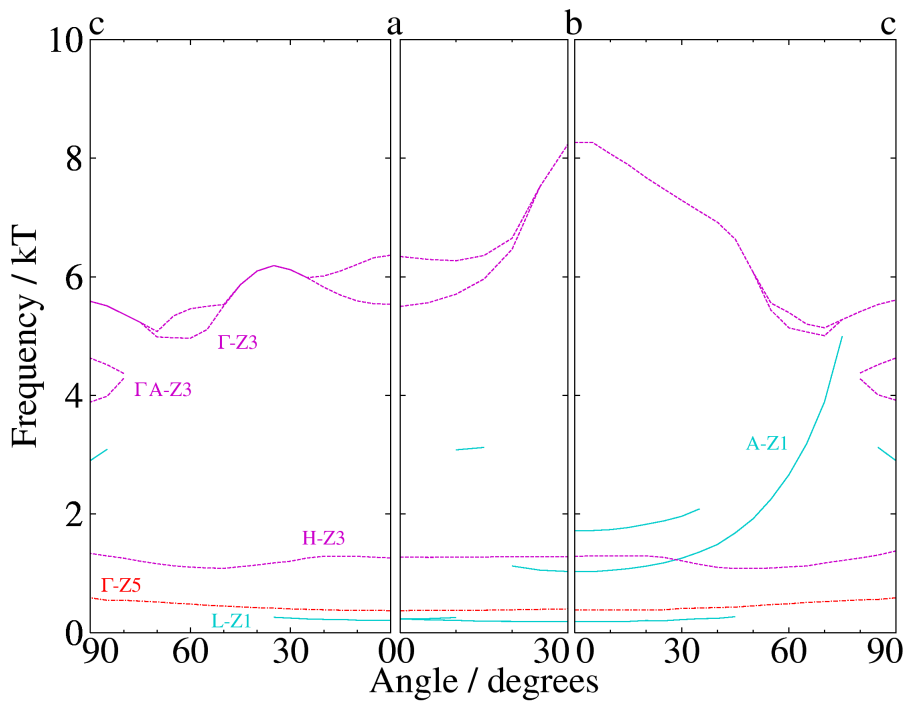}
\end{center}
\caption{
Predicted variation of the dHvA frequency components in
the \Zwicknagl\ model.
The label indicates the orbit centre and bands (Z1, Z3 and Z5).
}
\label{Zwicknagl_freqs}
\end{figure}

It is evident that 
there are considerable differences 
between the \Norman\ and \Zwicknagl\ 
Fermi surfaces, and the 
main  purpose of this  paper 
is 
to address the 
controversy over the number of valence 
f-electrons 
in UPt\(_3\) by comparing these model Fermi surfaces 
with experiment. 
In making this comparison, 
we make use of the Onsager relation, 
$F = \hbar A/2\pi e$, 
which relates an extremal cross-sectional area \(A\) of the 
Fermi surface 
to the measured dHvA frequency \(F\) \cite{Shoenberg_84}. 
Thus, 
in addition to figures \ref{valence_FS} and \ref{Zwicknagl_FS}, 
we show in figures  \ref{Freqs_vs_FS} and \ref{Zwicknagl_freqs} 
the variation with angle 
of 
the predicted dHvA oscillation frequencies 
for the \Norman\ model and \Zwicknagl\ models, respectively,  
extracted from band structures using the Onsager relation. 
In figure \ref{Freqs_vs_FS} we have included several predicted 
orbits on the band 2 surface of 
the \Norman\ model that have not previously been noted. 
These were obtained using a numerical method that 
interpolates the eigenvalues of the band structure calculation between $k$ points, locates 
closed orbits on the Fermi surface (including orbits spanning several Brillouin zones), 
identifies the extremal (maximum or minimum) orbits and calculates their areas.
Table \ref{tab-orbits} describes these orbits, and gives the correspondence with 
the orbits drawn on figures \ref{valence_FS} and \ref{band2_extended}. 

Most of these `new' orbits are topologically complicated, spanning two or more arms, 
often non-centrally. 
For example, AHK-2 circles two arms of band 2 
(yellow line in figure \ref{valence_FS}(b)) and 
there are two predicted AH-2 orbits that circle four arms of the Fermi surface 
(purple line in figure \ref{valence_FS}(b)).
A comparison of these orbits 
with several newly observed Fermi surface branches in our 
dHvA and Shubnikov--de Haas (sdH) measurements 
is given in the appendix.

\begin{table}[h]
\caption{
Orbits on the \Norman\ model (figure \ref{Freqs_vs_FS}), arranged roughly by 
increasing frequency.   The last column gives the proposed correpondence 
to observed branches of the Fermi surface (see e.g.\ figure \ref{Norman_angle})
}
\label{tab-orbits}
\begin{indented} \item[]
\begin{tabular}{|l|l|l|c|}
\hline
Label & Description & Correspondence with figure \ref{valence_FS} & Assignment 
\\ 
      &              &    figure \ref{valence_FS} &    
\\ \hline \hline                   
L-2 & (The lowest frequency) Electron orbit & Purple line on 1(b) & Possibly \(\gamma\) 
\\ \hline
ML-2 & Circles one arm on band 2 & Turquoise line on 1(b)  & \( \alpha \)  
\\ \hline
\( \Gamma \)-5 & Central orbit around band 5 sphere & Yellow line on 1(e) & \(\gamma'\) 
\\ \hline
AHK-2 & Circles 2 arms & Yellow line on 1(b) & \(\eta'\) \\ \hline
AH-2 & Circles 4 arms & Purple line on 1(b)  & \(\alpha_4'\) and \(\alpha_4\) \\ \hline
\(\Gamma\)-4 & Electron orbit & Red line on 1(d) & \(\epsilon\) \\ \hline
KH-2 & (At c-axis) Electron orbit & White line on 1(b) & \( \zeta \)  \\ \hline
K-3 & Electron orbit & Yellow line 1(c) & \(\kappa\) \\ \hline
KH-2 & Near 70\(^{\rm o}\) in $c$--$a$ plane &  Not shown & \(\alpha_3\) \\ \hline
A-1 & Hole orbit & Red line on 1(a) & \( \delta \) \\ \hline
\(\Gamma\)A-3 & Non-central electron orbit & Turquoise line on 1(c) & \(\sigma\) \\ \hline
ALM-2 &  Circles 3 upper arms and one lower & Not shown & \(\alpha_3\) \\ \hline
A-2 & At \(b\)-axis & Red line on 1(b) &  \( \lambda \)   \\ \hline
A-2 & At \(c\)-axis & Blue line on 1(b) & \( \lambda' \) \\ \hline
\(\Gamma\)-3 & Electron orbit  & Blue line 1(c) & \(\omega\) \\ \hline
L-2 & High frequency, $a$--$b$-plane & Red line, Figure 12 & ? \(\eta\) \\ \hline
L-2 & Higher frequency $a$--$b$-plane & Blue line, Figure 12 & \( \eta \) \\  \hline
\end{tabular}
\end{indented}
\end{table}

In the next section, we describe our experimental results, and then 
in the discussion we compare our results with the predictions of 
the \Norman\ and \Zwicknagl\ models.  

\section{Experimental Results}

Quantum oscillation measurements involve considerable technical challenges, 
especially in the realm of sample preparation because 
the oscillations are exponentially damped by impurity scattering. 

The sample used in our 
magnetoresistance measurements is a 50 $\mu$m $\times$ 160
$\mu$m $\times$ 4 mm single crystal whisker, produced by quenching a 
stoichiometric melt in ultra-high vacuum.  The whisker was annealed near its
melting temperature ($\sim 1700^{\rm o}$C) 
for a short time, then near 1400$^{\rm o}$C for several
hours, and finally overnight at 1200$^{\rm o}$C.  The residual resistivity
of this sample was 0.03 $\mu\Omega$\,cm, at least a factor of two better than
a whisker crystal in which Taillefer and Lonzarich had previously 
observed weak
quantum oscillations in the magnetoresistance \cite{Taillefer_unpub}, and
roughly 10 times better than that of a sample used by Kimura \etal\ \cite{Kimura95} in their
first study of the angular dependence of the magnetoresistance.  

Magnetoresistance measurements were made by a standard four-terminal 
technique, with the current along the $c$-axis, and the field
in the basal plane.  Typical currents were 1 to 10 $\mu$A.

The sample used in the dHvA 
oscillation measurements was grown in 
Grenoble by Czochralski pulling from a water cooled copper crucible, 
using high purity starting materials.  The sample was annealed for 
7 days at 950\(^{\rm o}\) C under UHV. 
With the current in the $a$-direction the RRR was 283 (not
extrapolated to $T = 0$), and 
\(T_c\) was 540~mK, with a width of 8~mK.

All 
measurements were carried out in a specially designed
cryomagnetic facility, which has an 18 tesla superconducting magnet,
modulation coils which can supply up to 0.02 T at 20 Hz with the magnet at 18 T,
a top-loading dilution refrigerator with a base temperature of 6 mK, and an
externally controlled rotation stage which allows the angle between the applied
field and the symmetry axes of the crystal to be varied.  The quantum 
oscillatory magnetization was measured in a standard way, with the sample
placed in one coil of a copper-wire astatic pair.   The modulation field was
typically 0.01 T at 2 Hz, the low frequency being used 
in order to minimize eddy current heating. 

The interpretation of quantum oscillations
in strongly correlated electron systems has recently  been
reviewed \cite{Shoenberg_84,Wasserman_96}. 
Briefly, thermodynamic and transport properties
oscillate as a function of applied field $B$ due to
passage through the Fermi surface
of quantized cyclotron orbits (called Landau levels) of charged
fermion quasiparticles.
Each measured
oscillation frequency
$F$ is related to an extremal Fermi surface area $A$ (measured perpendicular
to the applied field) by the Onsager relation
$F = \hbar A/2\pi e$.
Quantum oscillations are only seen if the width of the Fermi distribution is 
narrow compared with the Landau level spacing, 
so the oscillations are only seen at low
temperature: in the measurements described here the temperature 
was typically below 30 mK.

\begin{figure}
\begin{center}
\leavevmode
\epsfbox{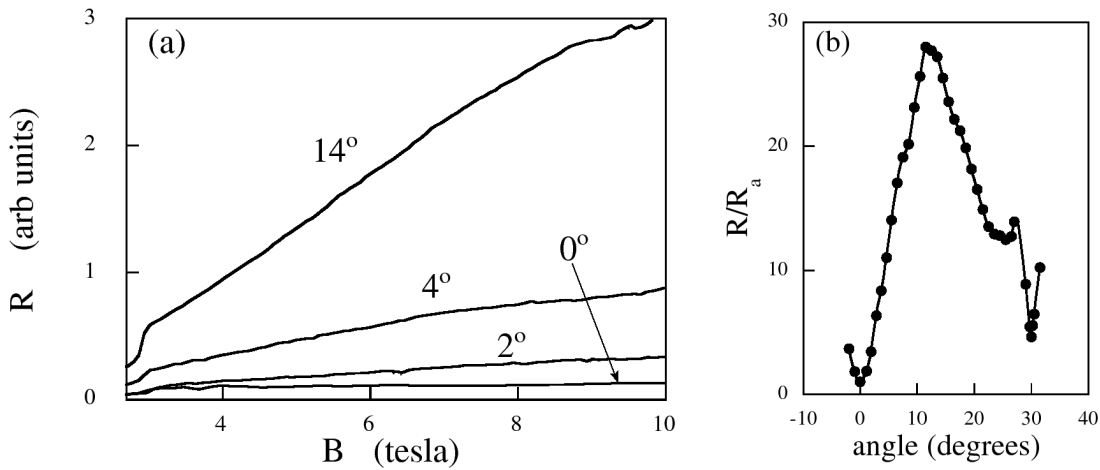}
\end{center}
\caption{(a) Resistivity versus field for various in-plane angles between the field and
the $a$-axis.  The very weak magnetoresistance when the field is aligned with the $a$-axis
($0^{\rm o}$ line) is strong evidence for an open orbit. 
(b) Resistivity versus angle at 8 T and 20 mK, showing the sharp
dips with the field along the $a$ (\(0^{\rm o}\)) and $b$ (\(30^{\rm o}\)) axes.} 
\label{R_vs_B_and_theta}
\end{figure}

Our results for the transverse dc-magnetoresistance (i.e.\ the non-oscillatory 
magnetoresistance) are illustrated in
figure \ref{R_vs_B_and_theta}, 
which shows the field dependence of the $c$-axis resistivity
at 20 mK for four different orientations of the applied field, and 
the variation of the
$c$-axis resistivity at $B=8$~T as the field is rotated within the 
basal plane. 
The strong angle dependence 
seen in our 
results 
contrasts sharply with 
\cite{Kimura95,Kimura96} where only
extremely weak anisotropy was found in the basal-plane magnetoresistance,  
but it is in excellent agreement with Taillefer \etal\ \cite{Taillefer_88b} and reference \cite{Kimura98}. 
Indeed, we see 
stronger angle dependence than in previous work, an 
indication of improved sample quality.  

\begin{figure}
\begin{center}
\leavevmode
\epsfbox{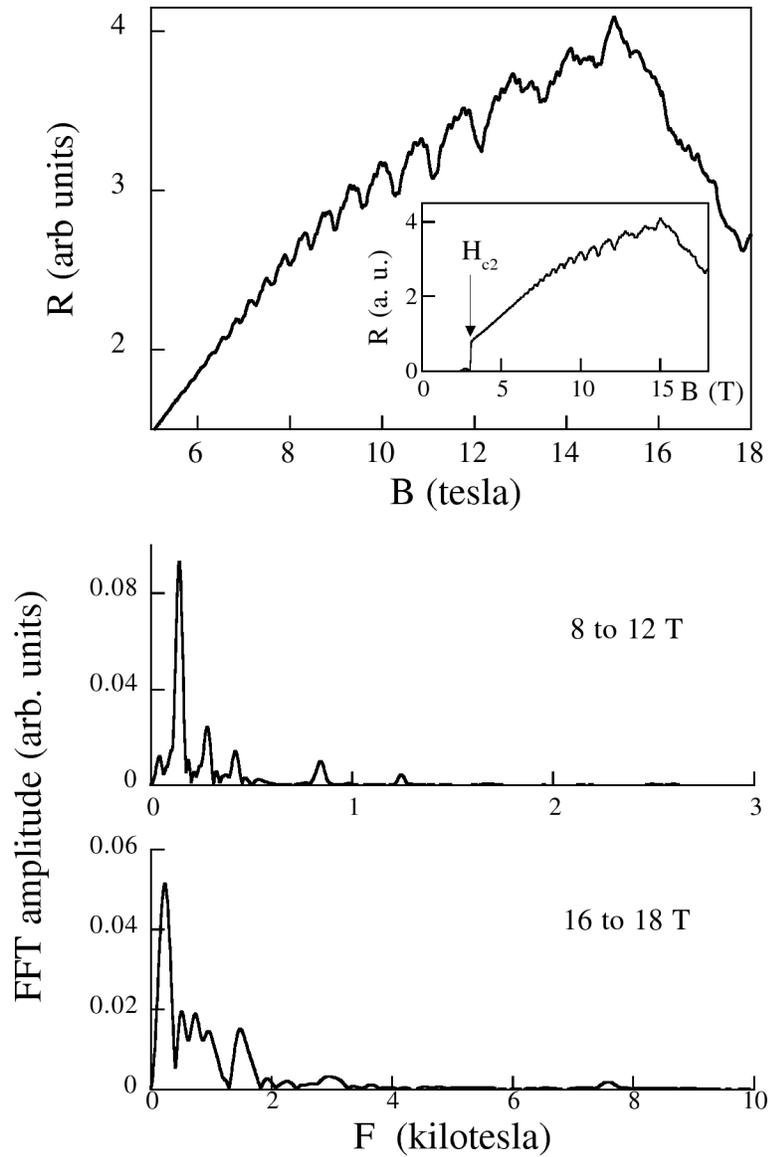}
\end{center}
\caption{The quantum oscillatory magnetoresistance in our high purity
whisker crystal of UPt$_3$, with the field applied in the basal plane
about 8$^{\circ}$ from the $b$-axis, and the current along $c$. The most
prominent oscillations correspond to a previously unobserved frequency.
The upper plot shows a complete trace in
the inset, taken at 20 mK, whereas the main plot focuses on the oscillations.
The middle plot is a quantum oscillation spectrum obtained from the data 
between 8 and 12 T by interpolating it at equally spaced intervals in \(1/B\) 
and taking the fast Fourier transform.  The largest peak, at 0.14 kT, is the new 
frequency, 
and it is labelled \( \gamma'\) in figures \ref{Norman_angle} and 
\ref{Zwicknagl_angle}.  
The bottom plot
is the quantum oscillation spectrum from the data between 16 and 18 T.
The peak at 7.5 kT is also a previously unobserved frequency, labelled 
$\eta$ in figures \ref{Norman_angle} and \ref{Zwicknagl_angle}.}
\label{Qm_osc_R}
\end{figure}

\begin{figure}
\begin{center}
\leavevmode
\epsfbox{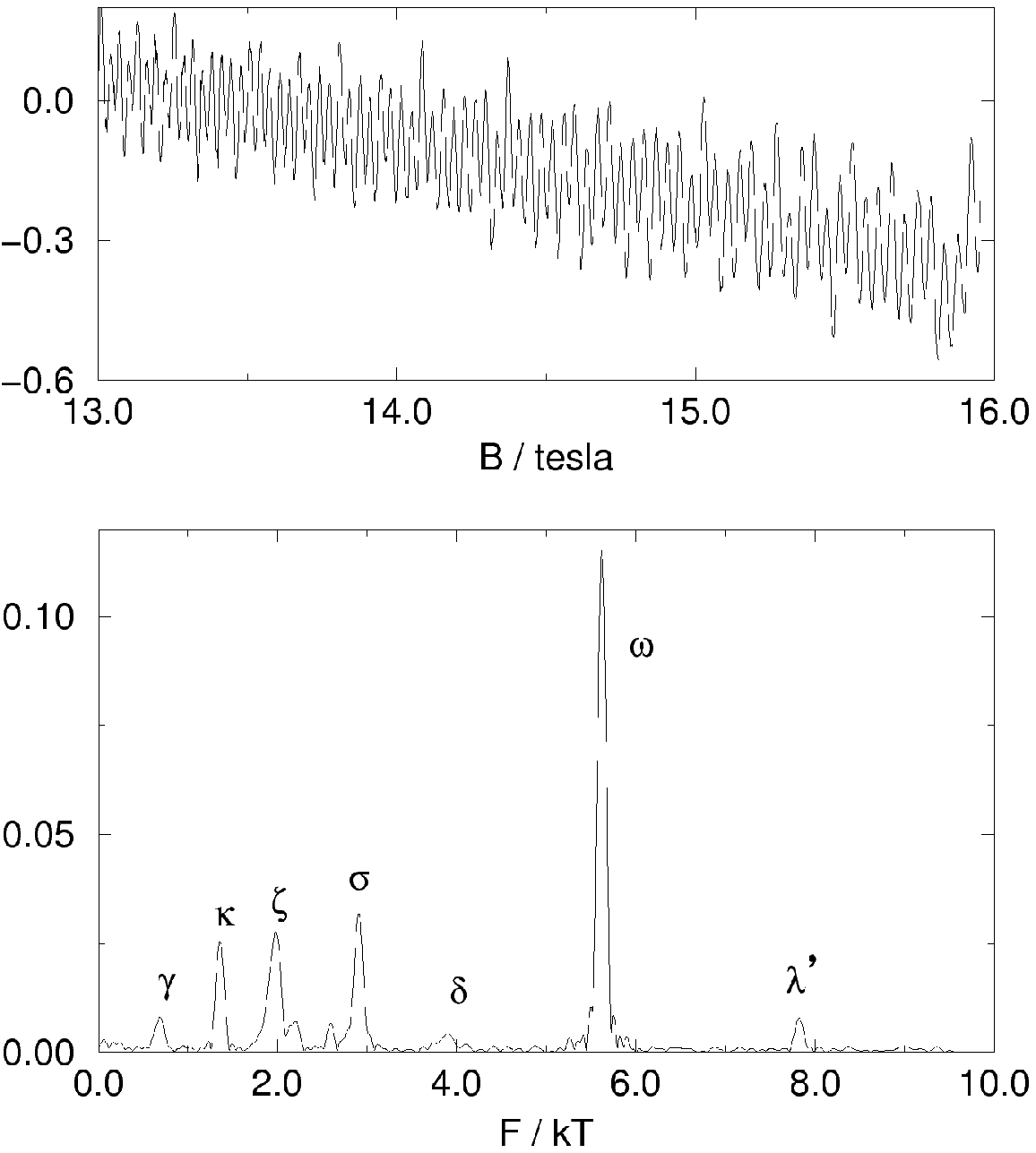}
\end{center}
\caption{Typical oscillatory variation of the dHvA magnetization as
seen at 10 mK with the applied field directed approximately 5$^{\circ}$ from the
$c$-axis towards the $b$-axis (upper trace) and corresponding Fourier
spectrum (lower trace).
The peaks are labelled according to our assigment on the rotation
plots (see figure \ref{Norman_angle}).  
}
\label{dHvA_nr_c}
\end{figure}

Quantum oscillations in the magnetoresistance of our single crystal
whisker, with the magnetic field applied in
the basal plane, are shown in figure  \ref{Qm_osc_R}, together with the
amplitude spectra over two field ranges: from 8 to 12 T, and 16 to 18 T.   
Figure \ref{dHvA_nr_c} shows typical quantum oscillatory magnetization 
for the Grenoble single crystal,
detected at the second harmonic of the modulation frequency, and taken 
with the magnetic field aligned close to the
$c$-axis.  

\begin{figure}
\begin{center}
\leavevmode
\epsfxsize 13.0 cm
\epsfbox{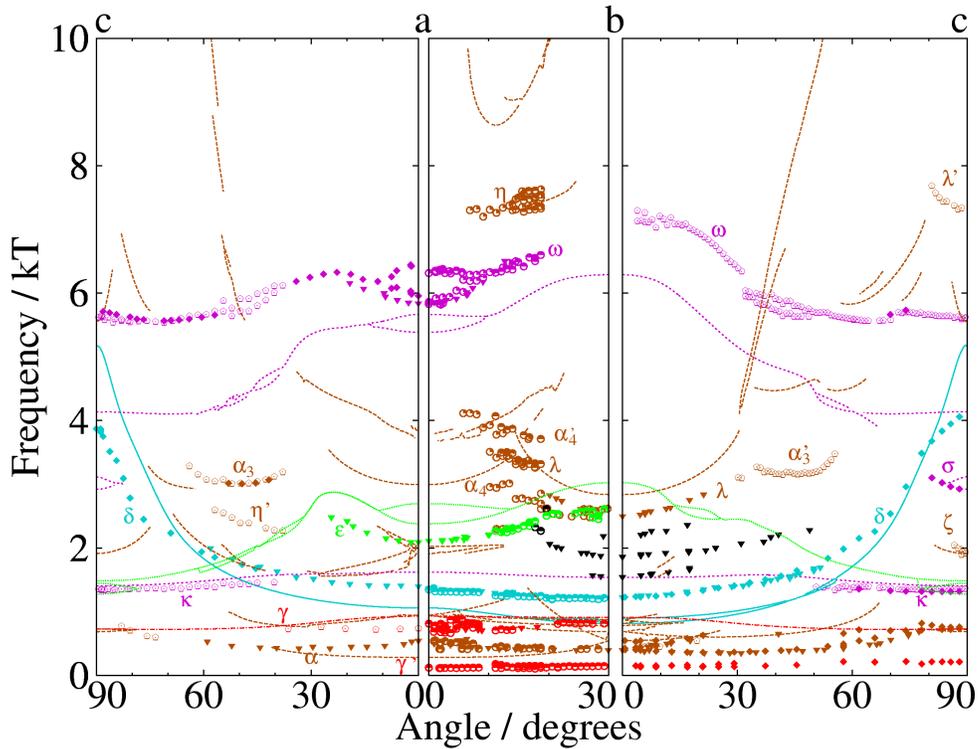}
\end{center}
\caption{ Angle dependence of measured quantum oscillation frequencies. 
The lines are the predictions of the \Norman\ model. The colour 
code reflects our band assignments as in figure \ref{valence_FS} (e.g.\ light 
blue is for frequencies arising from band 1, brown from band 2, etc). Black symbols are 
believed to be so-called breakdown orbits in which the quasiparticles tunnel between 
bands 1 and 2 as they undergo cyclotron motion \cite{Taillefer_88}. 
Triangles are the data of Taillefer and Lonzarich \cite{Taillefer_88}, 
circles are from our quantum oscillatory magnetoresistance measurements (e.g.\ 
figure \ref{Qm_osc_R}), filled diamonds are low-field (8--12 T) dHvA data, 
pentagons are from high-field (14--18 T) dHvA data.} 
\label{Norman_angle}
\end{figure}

Scans such as those shown in figures \ref{Qm_osc_R} and \ref{dHvA_nr_c}
were repeated at many angles, 
with 
the oscillatory magnetoresistance measurements 
focused on the 
$a$--$b$-plane, 
while the dHvA measurements were done only in the 
$a$--$c$- and $b$--$c$-planes, 
with somewhat stronger focus on the $b$--$c$-plane and 
with the crystals oriented so that signals were strongest 
near the \(c\)-axis.  
Field sweeps were typically divided into two ranges, 
from 8 to 12 T, and 14 to 18 T.   
The results are summarized in figure \ref{Norman_angle}, 
which shows the fundamental dHvA frequencies versus 
angle for the three major symmetry planes.  

The data  
of figure \ref{Norman_angle} 
have been `cleaned' to eliminate higher 
harmonic frequencies and occasional isolated
points. 
We have included the results of Taillefer and Lonzarich \cite{Taillefer_88}, 
which are in excellent agreement with ours, and which are 
complimentary in the sense that their dHvA 
studies of the $a$--$c$- and $b$--$c$-planes saw oscillations only 
at low angles (close to \(a\) and \(b\)), 
whereas our dHvA results are best close to the \(c\)-axis  
due to the configuration of our pickup coils. 
In comparing our results with the theory this  
limitation should be kept in mind, and it probably 
explains why, for example, we see branches that may correspond 
to AH-2 in the $a$--$b$-plane where the SdH 
oscillations were very strong, but we don't see them near the 
\( a \) axis in the $a$--$c$-plane, where our dHvA signals 
were comparatively weak. 

Like Kimura \etal\ \cite{Kimura95,Kimura96,Kimura98}, we were able to follow 
many branches to the $c$-axis, 
and in general our results are in good agreement with theirs, 
but in addition we observe a number of new frequencies. 

The Greek letters in figure \ref{Norman_angle} follow 
 \cite{Taillefer_88,Kimura98} 
with additions for new orbits. 
We have observed 
nine new orbits in all, which we have labelled 
\( \lambda',\ \alpha_3,\ \alpha_3', \ \alpha_4,\ \alpha_4',\ \kappa,\ \gamma',\ \zeta,\ \eta,\ \eta'\).
Athough effective masses are not the focus of this paper, in table \ref{mass_table} 
we give the masses, and the calculated band masses (assuming that the \Norman\ model 
is correct), for these new orbits. 

\begin{table}
\caption{
Measured effective mass and calculated band mass from the \Norman\ model 
for newly observed dHvA orbits.
}
\label{mass_table}
\begin{indented} \item[]
\begin{tabular}{|c|c|c|c|c|}
\hline
Orbit &  Plane &  \( m^*/m_e \) &  \( m_{band}/m_e \) & Fully itinerant model frequency \\ \hline
\( \kappa \) &   ($c$--$a$) and ($b$--$c$) & \(70 \pm 10\)  & 3.5  & K-3, at \(c\)-axis  \\ \hline 
\( \eta   \)&    ($a$--$b$) &       \(130\pm 20\)  & 13.0 & 
                L-2 (high frequency), 15\(^{\rm o}\) from \(a\) in $a$---$b$-plane \\ \hline 
\( \alpha'_4 \) & ($a$--$b$) &       \(110\pm 30\)  &  8.0  &
                 AH-2 (upper branch), 10\(^{\rm o}\) from \(a\) in $a$--$b$-plane\\ \hline
\( \alpha_4 \) & ($a$--$b$) &       \(80\pm  20\)  &  4.2   &
                 AH-2 (lower branch), 10\(^{\rm o}\) from \(a\) in $a$--$b$-plane\\ \hline
\( \lambda' \) & ($b$--$c$) &       \(70\pm   5\) &   4.8  & 
                 A-2, at \(c\)-axis \\ \hline
\( \alpha'_3 \) & ($b$--$c$) &      \(55\pm   5\) &   7.7  & 
                 ALM-2, 50\(^{\rm o}\) from \(b\) \\ \hline
\( \zeta \) & ($b$--$c$) &      \(70\pm  15\) &   2.8      & 
                 KH-2, at \(c\)-axis \\ \hline
\( \gamma' \) &    ($a$--$b$) and ($b$--$c$) &\(16\pm   3\)  &  1.9 & 
                 \(\Gamma\)-5, 10\(^{\rm o}\) from \(a\) in $a$--$b$-plane \\ \hline
\end{tabular}
\end{indented}
\end{table}

An additional important difference from previous studies is that 
we have followed the \( \delta \) orbit all the way to the c-axis, 
whereas previously it had only been followed to within about 
20$^{\circ}$ of \( c \). 
The significance of this is discussed below.  

\begin{figure}
\begin{center}
\leavevmode
\epsfxsize 13.0 cm
\epsfbox{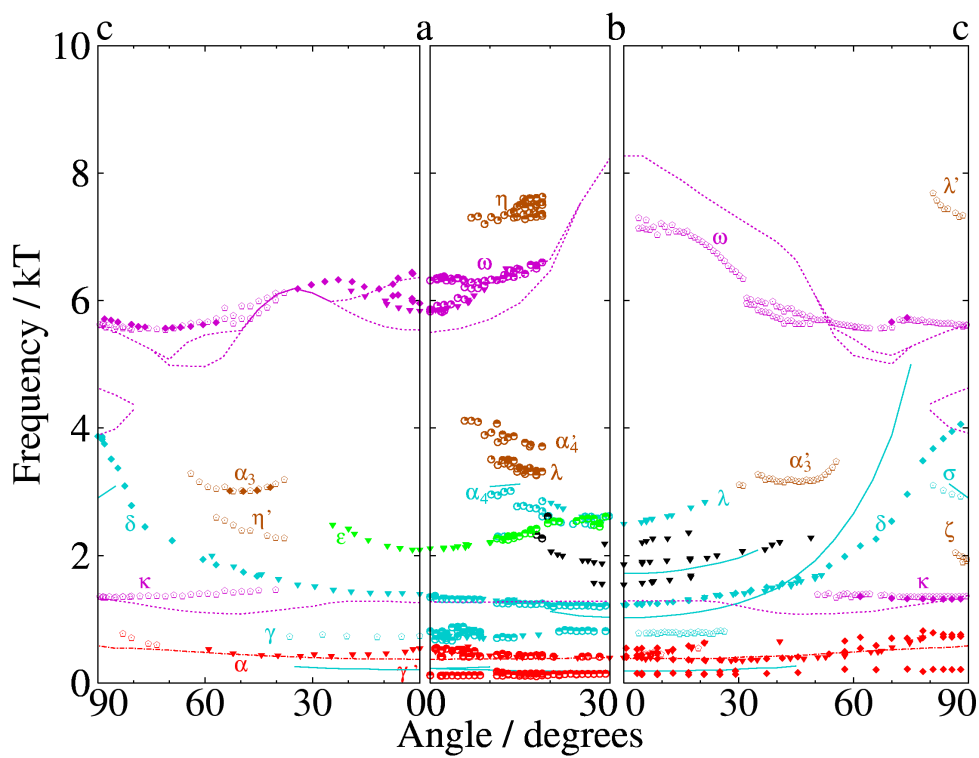}
\end{center}
\caption{ Comparison between measured quantum oscillation frequencies (same data 
as figure \ref{Norman_angle}) 
and the predictions of the \Zwicknagl\ model (solid lines).  }
\label{Zwicknagl_angle}
\end{figure}

The lines on figure \ref{Norman_angle} are the predictions of the \Norman\ model 
(as per figure \ref{Freqs_vs_FS}), and 
in the subsequent figure, figure \ref{Zwicknagl_angle}, 
we show the data together with the predictions of the \Zwicknagl\ model
(figure \ref{Zwicknagl_freqs}). 
In the next section, we carry out a detailed comparison between 
the data and the predictions of the models. 

\section{Discussion}

The extreme angle dependence of the dc-magnetoresistance, 
shown in figure \ref{R_vs_B_and_theta}, 
can be explained following Taillefer \etal
\cite{Taillefer_88b} who argue that it arises from
canonical $\omega_c \tau > 1$ effects.  They point out (a) that the large
magnetoresistance seen at most angles is indicative of an open orbit, since
UPt$_3$ is compensated and therefore ought not to have a magnetoresistance
unless there are open orbits, and (b) that the low magnetoresistance with the
field along the
$a$-axis is plausibly explained if the open orbit spans the arms of the 
band 2 surface, 
as shown in figure \ref{valence_FS}(b).
Within this explanation,
in \cite{Kimura95} 
a strongly anisotropic magnetoresistance was not seen 
because their measurement
was  made at too high a temperature and in a
lower quality
crystal,
so that the
$\omega_c\tau > 1$ condition was not reached.
(To observe an open orbit in the magnetoresistance, the 
quasiparticle mean free path must be 
at least 
long enough that quasiparticles, driven
by the Lorentz force, can
cross the Brillouin zone before they scatter.)
Similarly,
the reason we see a stronger magnetoresistance than Taillefer \etal
\cite{Taillefer_88b} is because our sample is purer, and so has a larger
value of $\omega_c\tau$.
The $\omega_c\tau$ interpretation is  further supported by Taillefer \etal \cite{Taillefer88a}, who show that the transverse magnetoresistance,
with $B$ lying between the $a$- and $b$-axes,
rises dramatically
only
at temperatures below 0.6 K,
that is, when the Fermi liquid thermal scattering rate falls below
\( \omega_c \),
as expected from an
$\omega_c \tau > 1$ effect.

We note that,
if the dip in the magnetoresistance
at $B\parallel a$ shows that an open orbit runs diagonally across the band A-2 surface
from one arm to  the other, then the dip at $b$ appears to suggest a
second open orbit,
possibly running non-centrally along the top of  the band 2 surface. 
We have also indicated this possible open orbit in 
figure \ref{valence_FS}(b).

Within the \Norman\ picture
there is thus excellent agreement between the
topology of the band 2 surface
and the angle dependence of the magnetoresistance.
However, the Z1 Fermi surface
in the \Zwicknagl\ model also has open orbits, very similar to those
of the band 2 surface of the \Norman\ model,
so the angle-dependent magnetoresistance
alone does not discriminate between
these models.
However, the existence of these open orbits must be kept in
mind in considering any
fine-tuning
of the \Zwicknagl\ model
that will be required to obtain agreement with the quantum oscillation
data.

Turning to the quantum oscillations,  we firstly note 
that there are major differences in the 
predictions of the \Norman\ and \Zwicknagl\ theories. 
At the coarsest level of comparison,
it is evident that
the \Zwicknagl\ model predicts many 
fewer dHvA frequencies than does the \Norman\ model.
This mostly reflects the absence in the \Zwicknagl\ model
of a surface corresponding to the band 2 `octopus',
as this surface generates many orbits
in the \Norman\ model.

It is at this coarsest level of comparison that the 
deficiencies of the \Zwicknagl\ model are perhaps most evident, 
because 
it simply predicts many 
fewer 
dHvA orbits 
than are actually observed.
The frequencies labelled 
\(\alpha\), \(\alpha_3\), \(\alpha_3'\), \(\alpha_4'\), \(\eta\), \(\eta'\), \(\zeta\), 
\( \lambda \), \( \lambda'\)  and \( \epsilon \) find no 
explanation in the \Zwicknagl\ model, 
nor do the 
three 
frequencies between 1.5 and 2.5 kT near the $b$-axis, 
ascribed in \cite{Taillefer_88} to 
breakdown orbits between the band 1 and band 2 surfaces 
of the \Norman\ model.
Also, 
the \(\delta\) oscillation, arising from band 1/Z1, is 
clearly observed in the data to extend across 
the $a$--$c$-plane, 
whereas in the \Zwicknagl\ model it should not appear in this plane at all,  
due to the open orbits on the Z1 surface. 

We can understand the nature of these deficiencies 
better via 
a more detailed level of quantitative, band-by-band comparison. 

Starting with band 3/Z3 -- the largest, most thermodynamically relevant, sheet of the
Fermi surface --    
as pointed out by Zwicknagl \etal \cite{Zwicknagl},
the Z3 surface is in
better quantitative agreement with
the \(\omega\) orbit 
than is the band 3 sheet of the \Norman\ model. 
It is on this basis that Zwicknagl \etal\ \cite{Zwicknagl} claimed that the \Zwicknagl\ model
is superior.
If the \Norman\ band 3 surface is to be brought into agreement 
with the data it has to be increased in size: 
the angle dependence follows that of the data quite well, however. 

Bands 3/Z3 both have,
as well as the large $\Gamma$-centred surfaces, 
smaller, roughly spherical sufaces 
of similar size, but located 
at different points of the Brillouin zone: 
in the \Norman\ model  
there are $K$-centred ellipsoids as shown in figure 
\ref{valence_FS}(c). 
The \Zwicknagl\ model in contrast has \(H\)-centred ellipsoids, 
but of almost exactly the same size. 
The branches which we have labelled $\kappa$ in figures \ref{Norman_angle} 
and \ref{Zwicknagl_angle} 
{\em could} arise from either of these surfaces 
(see the branches labelled K-3 and H-Z3 in figures \ref{Freqs_vs_FS} and \ref{Zwicknagl_freqs}).  
This is a very weak oscillation in our data, 
not previously observed. 
One issue with an assignment to K-3 is that increasing the
size of the band 3 $\Gamma$ surface as discussed in the previous paragraph would
also cause the $K$-centred pockets to increase, as they are also electron-like.
In fact, one problem with the \Norman\  model is that to increase the size of the $\Gamma$
surface to agree with experiment, one has the danger that this surface
merges with the $K$-centred ellipsoids.  
To correct for this would require a distortion in the dispersion
of this band.
Note, however, that, as discussed below, 
these  $\kappa$ oscillations could also arise from the band 4 surface of the \Norman\ model.

Turning next to band 1/Z1 (the $\delta$ branch) 
the \Zwicknagl\ model gives  
the best 
prediction of the quantitative value of this frequency
at the \(b\)-axis, 
however, only the \Norman\ model 
correctly predicts: (1) that this oscillation spans the entire $a$--$b$-plane, 
(ii) that it will be seen in the $a$--$c$-plane, 
and (iii) that it will extend to the \(c\)-axis in both the $a$--$c$- and 
the $b$--$c$-planes. 
These discrepancies between the \Zwicknagl\ model and the data 
were not addressed by Zwicknagl \etal \cite{Zwicknagl},  
yet it 
categorically shows that the topology of the Z1 surface is wrong: 
this surface is closed 
(i.e.\ it cannot have arms extending to the zone boundary) 
in contrast to figure \ref{Zwicknagl_FS}(a). 
Yet, as noted above, 
open orbits are required 
in order to explain the 
angle dependence of the magnetoresistance. 
Only by introducing an additional Fermi surface with open 
orbits, 
and adjusting the Z1 surface so that it is closed, 
can the \Zwicknagl\ model be brought into agreement with 
both the angle dependence of the $\delta$ oscillation and 
the angle dependence of the magnetoresistance. 
It is thus clear that not only is the Z1 topology wrong, 
there is also a major missing Fermi surface in the 
\Zwicknagl\ model. 

Band-2 of the \Norman\ model 
has open orbits, as shown in figure \ref{valence_FS}(b), 
that explain the angle dependence of the magnetoresistance. 
It 
was found in previous studies to provide a natural explanation of 
the \( \alpha \) and \( \lambda \) oscillations.
However, these find 
alternative explanations in the \Zwicknagl\ model: 
the \( \lambda\) oscillation follows the angle dependence of 
a non-central orbit on the Z1 surface, as illustrated in 
figure \ref{Zwicknagl_angle}, while \( \alpha \) could be 
explained by either 
\( \Gamma\)-Z5 or L-Z1.  
However, we believe that several of our new frequencies 
provide unambiguous evidence for the existence of the band 2 surface: 
the \( \lambda' \) orbit 
finds a natural interpretation  in the \Norman\ model as the 
orbit encircling the \lq waist' of this surface when the field is 
parallel to \( c \) (this is the blue line in figure \ref{valence_FS}(b)), 
although the frequency is 
slightly larger than that predicted. 
This very clear frequency (shown in figure \ref{dHvA_nr_c}) 
has not been previously reported. 
We found that it has a high mass: $70 \pm 5 m_e$. 
We note that Kimura \etal\ \cite{Kimura95} also believed that they had found 
this orbit, which they called \(\tau\), but it was at a lower frequency, and 
we did not observe this signal at all. 
We believe that our \( \lambda' \) frequency is in better agreement 
with the overall size of the band 2 surface. 
We note additionally that the 
\( \eta \), \( \eta'\), \(\alpha_3'\), \(\alpha_4\)  and \( \alpha_4' \)  
frequencies 
are only explainable via the topologically complicated 
orbits on the band 2 surface, and not within the \Zwicknagl\ 
model, as shown in figure 
\ref{Norman_explained}.  
This is discussed further in the appendix, 
but we draw attention particularly to the \( \eta \) orbit, 
a strong, high frequency oscillation in the basal plane, that corresponds 
well to non-central orbits that span two Brillouin zones, 
as illustrated in figure \ref{band2_extended}. 

Final, circumstantial evidence in favour of the existence 
of the band 2 surface 
is that it makes a major (roughly 30\%) contribution 
to the linear coefficient of specific heat, and if it does not exist then the 
linear coefficient of the specific heat 
is no longer explained by the quasiparticle 
masses.  

It thus seems very likely that it is this 
Fermi surface which is missing from the \Zwicknagl\ model. 

The $\epsilon$  orbit has previously been assigned to 
band 4 in the \Norman\ model, i.e.\ 
$\Gamma$-4 \cite{Taillefer_88},  
but it was not clear 
why this oscillation was only seen near the 
basal plane when it should be seen at all angles. 
A second possible interpretation of our 
\(\kappa \) oscillation is that it is the extension of this oscillation, 
but if this is true then the \( K \)-centred orbits of band 3 either 
do not exist, or their oscillations are even weaker than these signals.

Finally, turning to band 5, 
one of the mysteries of previous studies is why this small
electron sheet, with orbit 
$\Gamma$-5, does not show up more clearly.  
Taillefer and Lonzarich \cite{Taillefer_88} 
assigned it to a frequency of about 0.7 kT 
near the $a$-axis with the field in the basal plane.  
We see
a corresponding frequency extending toward the $c$-axis in the $a$--$c$-plane, 
within about 30$^{\circ}$ of the basal plane, but the signal then 
disappears abruptly, 
which is not easily explained for a simple spherical 
surface.  

A new candidate for this surface 
is, 
however, 
the strong low-frequency signal seen in the 
magnetoresistance (the most visible oscillation in figure \ref{Qm_osc_R}), 
with a frequency between 0.1 and 0.2 kT, 
labelled $\gamma'$ in figure \ref{Norman_angle}. 
Although it is very strong in the
magnetoresistance, it has not been previously reported, presumably because low
frequencies are much harder to see in dHvA. 
We  believe that the surface which gives rise to this
oscillation is roughly spherical,  because we were eventually 
able to follow it in dHvA measurements toward the $c$-axis  in the
$b$--$c$-plane, although we 
did not observe it along the $c$-axis, nor in the $a$--$c$-plane.   
The amplitude of this oscillation in the magnetoresistance 
is really huge (see figure \ref{Qm_osc_R}), given  the small size of
the Fermi surface, and it is also strongly angle-dependent 
The signal is
strongest when the field is in the basal plane about 10$^{\circ}$ from $b$:
in this angular region, the non-oscillatory magnetoresistance has a
shoulder (see figure \ref{R_vs_B_and_theta}).  
We have no explanation of the strength of this SdH oscillation; 
sometimes magnetoresistance
oscillations from small surfaces can be strong if the surface acts as an 
\lq interferometer' for larger Fermi surfaces \cite{PippardBook},  
but we do not have a model for such an interferometer in the Fermi 
surface shown in figure \ref{valence_FS}.
If indeed these oscillations arise from $\Gamma$-5, then
the surface in the \Norman\ calculation is too large. 
The quasiparticle mass on this surface was measured to be 
16 \(\pm\) 3 \(\rm m_e\). 

Finally, we note that 
the original \Norman\ model, 
while generally in much better agreement with the data, 
still has quantitative discrepancies with the data that are at least 
as striking as those that have recently aroused  interest in 
sodium cobaltate \cite{Yang05}.
Although it seems unlikely that better Fermi surface measurements 
will soon be on the horizon for UPt\(_3\), 
it may be that the present data set is sufficiently detailed 
that it could be used as the basis for 
improving electronic structure methods for 
f-electron systems.
For instance, it would be interesting if the Fermi surface of UPt\(_3\)
could be calculated from dynamical mean field theory, as this technique
has shown recent promise in other actinide materials such as 
plutonium \cite{Kotliar}.

\section{Conclusions} 

The Fermi surface of UPt\(_3\) 
has over the past two decades 
been experimentally determined to a high degree of accuracy, 
allowing comparison with the results of electronic structure 
calculations 
at both coarse and fine levels of detail.
Our new results fill some gaps in earlier data, 
and serve to emphasize the rather large number of Fermi surfaces 
of the heavy fermion state of UPt\(_3\), and the topological 
complexity of some of the surfaces.

We find that the \Zwicknagl\ model, 
which assumes that two 5f-electrons per uranium atom are localized, 
shows unacceptably large discrepancies with observations.
Although it gives good agreement with 
the large electron sheet, 
and some other oscillations, 
it has grave difficulty in explaining many of 
the observed oscillations.  
At the crudest level, there are a lot 
more oscillations present than can be supported on the 
\Zwicknagl\ Fermi surface.  
At a deeper level, the topology for the band Z1 surface 
does not agree at all well with the dHvA data, and if it is modified 
to agree, then one can no longer explain the angle dependence of the 
magnetoresistance. 
It is clear that a major, 
experimentally observed, 
sheet of the Fermi surface is missing from the \Zwicknagl\ model.
We have argued that the band 2 Fermi surface of the \Norman\ model 
must be close to the shape of this `missing' Fermi surface, and 
thus that the \Norman\ model of the Fermi surface provides a much 
better explanation of the data: thus we conclude that, 
on the basis of existing band-structure calculations, 
the experimental evidence very strongly favours the model in which all  
three of the uranium f-electrons are delocalized in UPt\(_3\).  

\section{Acknowlegement}

This work has been supported by the US DOE, Office of Science, under contract no.~DE-AC02-06CH11357,
the Natural Science and Engineering Research Council of Canada, the 
Canadian Institute for Advanced Research, and the Engineering and Physical Science Research 
Council of the UK. 

\section{Appendix. Assignment of band 2 orbits}

\begin{figure}
\begin{center}
\leavevmode
\epsfxsize 12.0 cm
\epsfbox{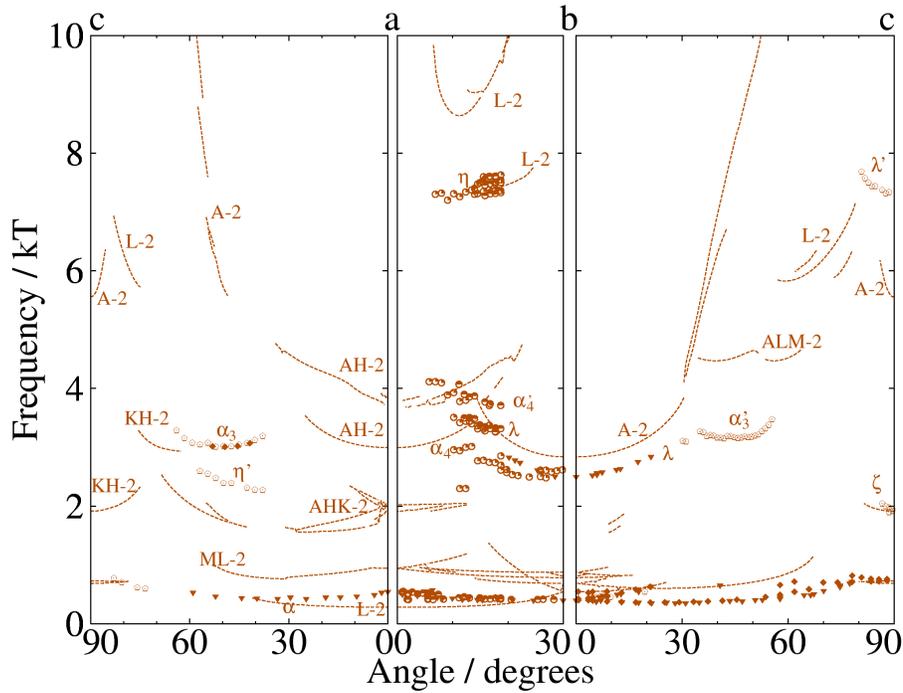} 
\end{center}
\caption{ Angle dependence of frequencies ascribed to band 2 in the \Norman\ model.
}
\label{Norman_explained}
\end{figure}

\begin{figure}
\begin{center}
\leavevmode
\epsfxsize 14.0 cm
\epsfbox{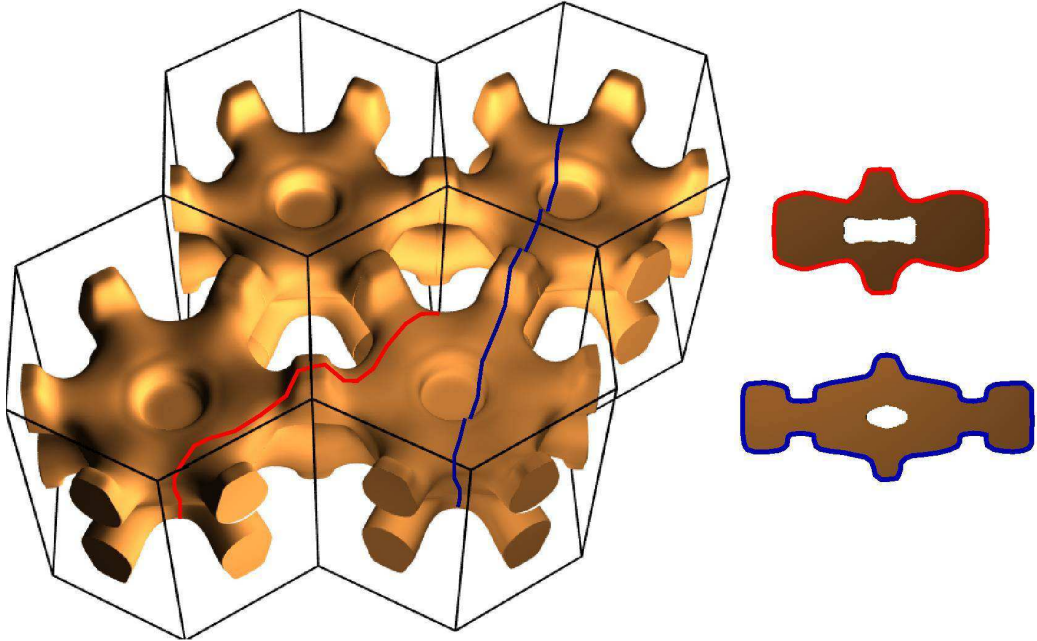} 
\end{center}
\caption{ The L-2 frequencies that lie between 6.0 and 10.0 kT between \(a \) and 
\(b\) span two Brillouin zones.  The main figure shows these orbits projected on the 
band 2 Fermi surface in the extented zone scheme, while the right shows the cross sections 
in cut-away view.  The lower of the two frequencies, outlined in 
red, is only marginally extremal: small shifts in the Fermi surface would 
eliminate this as an extremal orbit.  The higher frequency, shown in blue, is 
much more robust, extends over a greater range of angles, and is thus 
more likely to correspond to our \(\eta\) frequency.
}
\label{band2_extended}
\end{figure}

In this paper, a number of new quantum oscillation orbits are described, but 
we did not assign them to specific predicted orbits in the 
\Norman\ model because the case for preferring that model to the 
\Zwicknagl\ model can be made without resorting details. 
Nevertheless, it is possible to explain our new orbits using the 
band 2 surface (with the exception of \(\gamma'\), which we have ascribed to 
the \(\Gamma\)-5 orbit). 

Figure \ref{Norman_explained} shows the angle dependence of dHvA frequencies 
for the band 2 surface (solid lines) from the \Norman\ model compared with 
observed oscillation frequencies that we have ascribed to this surface.  

Recall that our two labelling systems are: 
for the theoretical prediction, 
letters and numbers  identify the centre of the orbit in \(k\)-space
and the band on which the orbit exists; 
for the experimental observation we use   
lowercase Greek letters. 

The following assignments had been made previously:  
\(\lambda\) is assigned to A-2, the orbit that circles the body of band 2 
(red line in figure \ref{valence_FS}(b));
\( \alpha\) is assigned to ML-2, the orbit that encircles one arm of band 2 
(turquoise line in figure \ref{valence_FS}(b)) \cite{Taillefer_88}.

Our best assignment of the `new' orbits is:
\(\alpha_4\) and \(\alpha_4'\) correspond to the two AH-2 orbits, 
which both circle four arms but at different 
distances out from the centre of the surface (purple line in 
figure \ref{valence_FS}(b)); 
\( \eta \) seems to agree well with L-2, which is the 
red orbit in figure \ref{band2_extended} which spans two zones, 
however, this orbit is only marginally extremal, so it might be 
better to ascribe this frequency to the larger, but more stable, orbit 
shown by the blue line on figure \ref{band2_extended} -- this orbit is 
also labelled L-2 and it too spans two Brillouin zones;
\( \alpha_3\) corresponds well to KH-2, while \( \alpha_3' \) 
corresponds (not quite so well) to ALM-2, both of which 
circle three upper arms and one lower arm when the field is at a rather high angle 
going towards \(c\) 
(these orbits are not 
shown on figure \ref{valence_FS}(b)); 
\(\eta'\) corresponds well to AKH-2, 
the orbit that circles two arms (yellow line on 
figure \ref{valence_FS}(b));
\( \zeta \) corresponds well to KH-2, which is an electron 
orbit that runs around the interior of the arms of three 
adjacent Brillouin zones (white line in figure \ref{valence_FS}(b)); 
and we believe that \( \lambda'\) is, as discussed in the text, 
the central orbit that runs around the waist of band 2 
(blue line in figure \ref{valence_FS}(b)).

The comparison between the newly observed orbits and the 
predictions for this surface is, we believe, very convincing, 
keeping in mind that there were some limitations to 
our experiment:  our dHvA signals were comparatively weak 
near the $a$--$b$-plane because of the orientation 
of the crystal in the pickup coils. 
Thus, the failure to see the AH-2 and the AHK-2 orbits 
near the \( a\)-axis in the $a$--$c$-plane 
is understandable, if unfortunate. 
Moreover, the quantum oscillatory magnetoresistance, which 
gave rise to the oscillations between 3.0 and 4.0 kT in 
the $a$--$b$-plane, lost intensity near the \( a \)-axis due to the near-vanishing of the magnetoresistance 
there; hence these oscillations could not be followed to 
the \( a \)-axis in the $a$--$b$-plane either.

Thus, the best regions for comparison are close to the 
\(c\)-axis in the $a$--$c$- and $b$--$c$-planes, 
and at intermediate angles in the $a$--$b$-plane, 
and this is precisely where we find many oscillations 
that are only explainable by band 2 of the \Norman\ model.

\end{document}